# Resource Orchestration of 5G Transport Networks for Vertical Industries

K. Antevski[1], J. Martín-Pérez[1,2], Nuria Molner[1,2]
[1]Universidad Carlos III de Madrid, Spain
[2]IMDEA Networks Institute, Spain

C. F. Chiasserini, F. Malandrino
Politecnico di Torino, Italy

P. Frangoudis, A. Ksentini
EURECOM, France

X. Li, J. SalvatLozano
NEC Laboratories Europe GmbH, Germany

R. Martínez, I. Pascual,
J. Mangues-Bafalluy, J. Baranda
Centre Tecnològic de Telecomunicacions de Catalunya
Castelldefels, Spain

B. Martini, M. Gharbaoui
Scuola Superiore Sant'Anna, Italy

*Abstract*—The future 5G transport networks are envisioned to support a variety of vertical services through network slicing and efficient orchestration over multiple administrative domains. In this paper, we propose an orchestrator architecture to support vertical services to meet their diverse resource and service requirements. We then present a system model for resource orchestration of transport networks as well as low-complexity algorithms that aim at minimizing service deployment cost and/or service latency. Importantly, the proposed model can work with any level of abstractions exposed by the underlying network or the federated domains depending on their representation of resources.

## I. Introduction

5G transport networks are envisioned to expand the service scope of current mobile networks to support various vertical services, such as eHealth, automotive, media, or cloud robotics, hence enriching the telecom network ecosystem. To enable such vision, the EU H2020 5G-PPP phase 2 5G-TRANSFORMER project [1] proposes a flexible and adaptable SDN/NFV-based transport and computing platform, capable of simultaneously supporting the needs of different vertical industries to meet their diverse range of resource and service requirements. In this design, Network Function Virtualization (NFV) and Network Slicing are the key solutions to address this challenge.

The 5G-TRANSFORMER solution consists of three novel building blocks, as illustrated in Figure 1:

1) **The Vertical Slicer (5GT-VS)** is the common entry point for all verticals and MVNOs (Mobile Virtual Network Operators) into the system. It dynamically creates and maps the vertical services onto network slices according to their requirements, and it manages their lifecycle. It also translates the vertical and slicing request into an NFV Network Service (NFV-NS) and sends it to the 5GT-SO, where a slice will be deployed as an NFV-NS instance.

2) **The Service Orchestrator (5GT-SO)** offers service or resource orchestration and federation. Orchestration entails managing end-to-end services or resources that may be split into multiple segments belonging to different administrative domains based on requirements and availability. Federation entails managing administrative relations at the interface between the 5GT-SOs of different domains and handling abstraction of services and resources.

3) **The Mobile Transport and Computing Platform (5GT-MTP)** is the underlying unified transport (and computing) stratum, responsible for providing the resources required by the NFV-NSs orchestrated by the SO. This includes their instantiation over the underlying physical transport network, computing, and storage infrastructure. It also needs to (de)abstract the 5GT-MTP resources offered to the 5GT-SO.

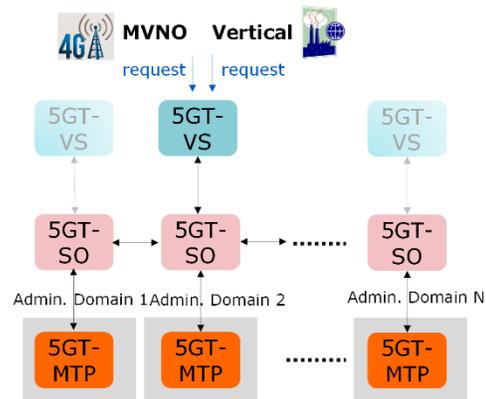

**Figure 1:** The 5G-TRANSFORMER concept

This paper focuses on the orchestration layer of the 5G-TRANSFORMER system and on how the 5GT-SO orchestrates resources across one or multiple administrative domains in order to deploy the requested NFV-NS. The orchestration decisions are based on the slice requirements expressed by the different verticals in their service request, which are, in turn, mapped to an NFV-NS request by the 5GT-VS. Network context (e.g., topology, available resources) is also taken into account. These decisions imply not only the allocation of the underlying network, computing and storage resources, and placement of virtual network functions (VNFs), but also the interaction (federation) with other administrative domains when, for instance, requirements cannot be met with services and resources of a single domain. In this way, the virtual resources offered by multiple infrastructure providers can be aggregated by federating them through their respective 5GT-SOs (see Figure 1).

We remark that resource orchestration is an extremely relevant problem, targeted by several works in the literature. Some, including [2][3], tackle the problems of VNF placement and routing from a network-centric viewpoint, i.e., they aim at minimizing the load of network resources. In particular, [2] seeks to balance the load on links and servers, while [3] studies how to optimize routing to minimize network utilization. The above approaches formulate mixed-integer linear programming (MILP) problems and propose heuristic strategies to solve them.

Other works take the viewpoint of a service provider, supporting multiple services that require different, yet overlapping, sets of VNFs, and seek to maximize its revenue. The work in [4] aims at minimizing the energy consumption resulting from VNF placement decisions. [5] instead studies how to place VNFs between network-based and cloud servers so as to minimize the cost, and [6] studies how to design the VNF graphs themselves, in order to adapt to the network topology.

Among more recent works, [7] addresses the VNF placement problem in a setting where the objective is to minimize service delay, and the assignment of computational resources to individual VNFs is flexible and impacts their service times. [8] targets scenarios where hosts are distributed across multiple, interconnected datacenters, and orchestration decisions must be made accounting (also) for the latency of inter-datacenter links. Finally, [9] targets the related problem of service composition, arising in scenarios where multiple services whose VNF graphs overlap have to be served by the same set of datacenters.

Our study differs from previous work since our goal is threefold: (1) to enable vertical industries to meet their specific service requirements through an efficient resource orchestration; (2) to expose capabilities of the underlying infrastructure via different levels of abstraction to the orchestration layer; (3) to aggregate and federate transport networking and computing fabric, from the edge up to the core and cloud, to create and manage slices throughout a federated virtualized infrastructure.

## II. 5G-TRANSFORMER SERVICE ORCHESTRATOR

Here we better detail the main tasks of the 5GT-SO, where our resource orchestration mechanisms will be implemented. As mentioned, the 5GT-SO determines resource allocation for the requested NFV-NSs and the placement of the associated VNFs over the 5GT-MTP(s). Additionally, it handles the operations required to deploy them and manage their entire lifecycle. Figure 2 presents the 5GT-SO functional architecture with a high-level overview of the main functional modules and the interactions that need to be developed to realize the orchestration operation.

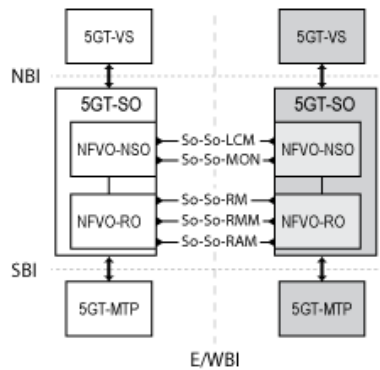

**Figure 2**: **5GT-SO functional architecture**

The main orchestration operations are handled using the NFVO Network Service Orchestrator (NFVO-NSO) and NFVO Resource Orchestrator (NFVO-RO) blocks. The NFVO-NSO is in charge of creating and deploying end-to-end network services as well as managing its entire lifecycle. In detail, the NFVO-NSO performs operations such as service on-boarding, instantiation, scaling, termination and management of the NFV Network Service, which is described by the so-called VNF Forwarding Graph (VNFFG) and associated deployment flavours [10]. Instead, the NFVO-RO decides the allocation of a set of virtual resources for each NFV-NS segment, and the placement of each involved VNF over the virtual infrastructure (either local or federated domains).

The 5GT-SOs belonging to different administrative domains interact with each other by using the So-So interface, defined as Eastbound/Westbound Interface (E/WBI). The E/WBI interface is used for enabling the service and resource federation between different administrative domains. The federation of services and federation of resources are two separate procedures that take place in different phases of the service instantiation/modification within the 5GT-SO. Service federation allows an administrative domain to request services that are instantiated and managed by other peering administrative domains. Resource federation, instead, allows an administrative domain to request, use, and manage resources that are owned by other peering administrative domains. As shown in Figure 2, out of the five reference points on the E/WBI, two are used for service federation and three for resource federation.

The decision about service federation is done during service instantiation/modification. Then, the 5GT-SO (consumer) sends an instantiation request to a peering 5GT-SO (provider), specifying the NFV-NS segment to be deployed in the peering domain. This request is sent through the So-So-LCM reference point of the E/WBI. The peering 5GT-SO can approve or reject the request (e.g., based on both service and resource availability). In case of acceptance, the peering 5GT-SO becomes the provider of that service segment (i.e., it is responsible for orchestrating the service segment in its own domain) and will send monitoring information back to the consumer 5GT-SO via the So-So-MON reference point. Near the end of the service instantiation operation, the consumer 5GT-SO completes the end-to-end service by linking the "consumed" federated segment with the others, through connection points.

Next, let us consider that a 5GT-SO has to instantiate a service, or a segment of it. Then the 5GT-SO needs to decide *which resources* should be used to that end. Note that, in case of resource shortage in its own administrative domain, the 5GT-SO may be entitled to use resources in other domains (*resource federation*), possibly, at a higher cost. In order to implement resource federation, the 5GT-SOs bi-directionally exchange information on the resources they have and that are available for federation, using the So-So-RAM reference point of the E/WBI. At each 5GT-SO, the resources available in other domains are stored into a database, which is kept up-to-date thanks to the dynamic repetition of the above information exchange. Upon any decision for resource federation, the requests for consuming federated resources are sent through the So-So-RM reference point of the E/WBI. Unlike in service federation, the consumer 5GT-SO has full access, management, and control of the (potentially abstracted) federated resources in the provider domain (through the So-So-RM reference point) in addition to their monitoring information (through the So-So-RMM reference point) and connection points.

## III. RESOURCE ABSTRACTION

In the 5G-TRANSFORMER architecture, the 5GT-MTP is responsible for providing the 5GT-SO with the information about the available resources, so that the 5GT-SO can make decisions on service instantiation. Because of the varying level of trust among organizations and the complexity associated to resource management, the 5GT-MTP, in general, does not provide all of its infrastructure details. Rather, it presents to the 5GT-SO the information with a certain *level of abstraction*. (For similar reasons, provider domains in resource federation may also abstract resources.)

Specifically, the resources controlled by an 5GT-MTP can be divided in two groups: computing resources and network transport resources. Computing resources are the physical machines that can accommodate VNFs and are typically characterized by CPU, memory and storage capabilities. Computing resources are grouped by location in NFVI Points of Presence (NFVI-PoPs), and the physical machines of an NFVI-PoP are managed by the so-called VIM (Virtual Infrastructure Manager), i.e., the software entity that actually manages (and reports on) the computing resources. Transport resources are represented by the network forwarding units and the physical links interconnecting them. WIMs (WAN Infrastructure Managers) are the entities that control network resources, also reporting the network topology and the available link bandwidth and latency.

An infrastructure can thus be represented as a composition of network and computing resources controlled by WIMs and VIMs, respectively. Since the nature of these resources is intrinsically different, the abstraction mechanisms for these two types of resources can also be different and can be combined as follows.

*Level 1:* also named WIM level because only WIM resources are abstracted. The 5GT-MTP reports all details about computing resources while the network resources are abstracted as a set of virtual links connecting the physical machines, with each link being characterized by a given bandwidth and latency.

*Level 2:* also named VIM level because, besides the WIM abstraction of level 1, the computing resources are aggregated per VIM. The 5GT-MTP reports the computing capabilities, CPUs, memory, storage, with an NFVI-PoP granularity instead of by physical-machine granularity as in Level 1. Regarding the network resources, only the connections between NFVI-PoPs are reported, as virtual links with a given bandwidth and latency.

*Level 3:* also named MTP level because all resources, both computing and network resources, are aggregated with 5GT-MTP granularity. This level may be useful for resource federation, as it allows a 5GT-SO to expose to peer 5GT-SOs the resources available within its administrative domain while hiding the complexity and the infrastructure details. In general, this higher level of abstraction is handled by the 5GT-SO, as it is the one to decide which levels of abstraction to be exposed to other 5GT-SOs, due to administrative or agreement on information constraints.

Our algorithms can conveniently work with any of the above levels of abstractions. We also remark that the selected abstraction levels between 5GT-MTP and 5GT-SO, and between the peer 5GT-SOs, may be different.

## IV. RESOURCE ORCHESTRATION ALGORITHMS

Below, we start by introducing the model, along with the variables and the constraints that characterize our system (Sec. IV.A). As shown in [7], the problem of resource orchestration in SDN/NFV system is NP-hard, which makes an optimal solution impractical in real-world conditions. Thus, we leverage on a heuristic approach and propose three swift, yet efficient, resource orchestration algorithms (Sec. IV.B).

*A. System model*

We consider that the 5GT-SO receives two main pieces of information, on which it can leverage to make orchestration decisions. The former is provided by the 5GT-VS and is given in the form of the service VNF Forwarding Graph (VNFFG), i.e., the set of VNFs and edges connecting them, and the deployment flavours, representing the service to be deployed and the associated requirements. The latter is provided by the 5GT-MTP and refers to the available resources. As discussed above, the representation of the resources depends on the abstraction that is used, however our algorithms can work with any level of abstraction. Thus, in the following we will refer to the resource representation as a host graph, where hosts (i.e., vertices) can be either physical machines, as per Level 1, or NFVI-PoPs, as per Level 2, and edges are virtual links (VLs) connecting hosts.

As far as the **VNFFG** is concerned, we denote its VNFs (i.e., vertices) by $v \in V$, each requiring an amount $r(v,\rho) \geq 0$ of resource type $\rho \in R$. Elements of the resource type set $R$ can include CPU, memory, and storage. $r$-values account for both the quantity of traffic each VNF has to process (e.g., in Mbits), and the amount of computational resources needed to process each unit of traffic (e.g., in CPU cycles per Mbit). Each time a request traverses a VNF, it incurs a delay $d(v)$. For each pair of VNFs $v_1, v_2 \in V$, i.e., for each edge of the VNFFG, we know the amount of traffic $f(v_1, v_2) \geq 0$ flowing from $v_1$ to $v_2$. Clearly, $f(v_1, v_2) = 0$ means that there is no traffic between those VNFs.

The 5GT-SO also knows the set of **services** to be deployed, $S = \{s\}$, the number of times $n(s, v)$ requests of service $s$ visit VNF $v$, the probabilities $P(v_2|v_1, s)$ that they visit $v_2$ immediately after $v_1$, and the maximum acceptable delay for that service $D^{max}(s)$.

The **host graph** has hosts $h \in H$ as vertices, each with capabilities $C(h, \rho) > 0$ for each resource type. Links (VLs) between hosts have a capacity $T(h_1, h_2)$, expressing the maximum *total* quantity of traffic that can flow per second from VNFs hosted at $h_1$ to VNFs hosted at $h_2$. Similarly, requests traveling a link incur a delay $\delta(h_1, h_2)$.

The main **decision** to make at the 5GT-SO is whether to place an instance of VNF $v$ at host $h$, expressed through a binary variable $x(h, v) \in \{0,1\}$. Each VNF placement incurs a cost $\kappa(h, v, op)$. A very relevant factor contributing to $\kappa$ is represented by the fees charged by different mobile operators, $op$, for the usage of their infrastructure by placing VNF $v$ at host $h$. The fees are pre-determined and defined by each operator. The maximum cost per service is denoted by $\kappa^{max}(s)$.

Two **constraints** must be honored, concerning the capabilities of hosts and the capacity of links, i.e.,

$$\sum_{v \in V} r(v, \rho) x(h, v) \leq C(h, \rho), \quad \forall h \in H, \rho \in R,$$

$$\sum_{v_1, v_2 \in V} x(h_1, v_1) x(h_2, v_2) f(v_1, v_2) \leq T(h_1, h_2),$$

$$\forall h_1, h_2 \in H.$$

Also, delay constraints, accounting for both processing and propagation delays, and cost constraints have to be met. For each service $s \in S$, the following must hold:

$$\sum_{v \in V} n(s, v) d(v) + \sum_{v_1, v_2 \in V} n(s, v_1) P(v_2|v_1, s) \cdot$$

$$\sum_{h_1, h_2 \in H} x(h_1, v_1) x(h_2, v_2) \delta(h_1, h_2) \leq D^{max}(s).$$

$$\sum_{v \in V, h \in H, op} k(h, v, op) x(h, v) n(s, v) \leq k^{max}(s).$$

*B. Minimizing service deployment cost and/or service latency*

We now introduce three heuristics that aim at minimizing the service cost and/or the service latency, while fulfilling all of the above constraints.

*1) Cluster-based approach*

The high-level goal of the cluster-based approach is to find the best tradeoff between the cost for the operator, as expressed by the $\kappa$-parameters, and the service latency. Our strategy is to take care of delay constraints and cost separately, in two different stages:

- first, we divide both the VNFFG and the host graph into *clusters* in such a way to guarantee low network delays;
- then, we assign VNFs in each VNFFG cluster to hosts in the corresponding host cluster so as to ensure low costs $\kappa(h, v, op)$.

**Clustering stage**. The intuition behind this stage is that, in order to meet service delay constraints, we must keep network delays low, and this in turn means having as little traffic as possible flowing on high-delay links between hosts. Therefore, we *cluster* both the VNFFG and the host graph in the same number of clusters, ensuring that: (i) in the VNF graph, high-traffic edges connect VNFs of the same cluster and low-traffic edges connect VNFs of different clusters; (ii) in the host graph, low-delay links connect hosts of the same cluster and high-delay links connect hosts of different clusters.

We adopt an iterative, hierarchical clustering technique, presented in [11] and implemented in [12]: at the first iteration, each node starts in its own cluster (*singleton*). At subsequent iterations, the two clusters connected by the highest-traffic edge in the VNFFG, and the two connected by the lowest-delay edge in the host graph, are joined together.

**Assignment stage**. In this stage, we have to decide at *which host* each VNF shall run. Thanks to the previous clustering stage, network delays can be ignored, while processing delays $d$ only depends on the VNF and not on the host at which it runs. Therefore, we can assign VNFs to hosts with the sole purpose of minimizing costs—specifically, we start from the VNF with the largest delay and place it at the cheapest host with enough spare resources to run it.

In many situations, multiple hosts will be associated with the same cost. In these cases, we break ties by trying to *balance* the load across different hosts. Formally, we choose the VNF to place and the host at which it should be placed so that the following quantity is minimized:

$$\max_{h \in H} \max_{\rho \in R} \frac{\sum_{v \in V} x(h,v) r(\rho,v)}{C(h,\rho)}.$$

In the expression above, the fraction represents how close to exhaustion resource $\rho$ is at host $h$. We seek to minimize the maximum of such ratios among all resources and hosts, thus reducing the risk to have, e.g., hosts with plenty of spare CPU but no free memory.

Importantly, each step of our approach has polynomial time complexity in the number of VNFs, hosts, and links; therefore, the global complexity is polynomial as well.

Furthermore, the approach can be easily extended to *multi-domain scenarios* where federation can be exploited in case of lack of resources in the domain controlled by the 5GT-SO that is in charge of deploying the service requested by the 5GT-VS. In particular, the algorithm can be extended as follows:
- in the clustering stage, edges connecting hosts belonging to different domains should be assigned higher weight, so as to limit the amount of traffic flowing across different domains;
- in the assignment stage, hosts belonging to foreign domains should be assigned higher costs, so as to model the fact that resources from foreign domains ought to be used only when necessary.

*2) Minimum-distance approach.*

This strategy aims at minimizing the consumption of network resources as well as the network latency (i.e., propagation delay) experienced by data while traversing VLs. In particular, the propagation delay is considered when data traverses the distance between network nodes connected to hosts. For ease of presentation, we describe the strategy by considering a VNFFG composed of two VNFs to be placed into as many hosts. The algorithm seeks for the pair of hosts with the shortest distance provided that the network path connecting them fulfills the bandwidth demand, the candidate hosts have sufficient available resources to meet CPU, memory and storage demand, and the candidate pair of hosts and network path satisfies the overall latency constraint. This strategy tends to consolidate utilizations in terms of both network and hosts resources at the cost of not achieving the lowest overall latency performance.

*3) Minimum-latency approach.*

We now aim at minimizing the overall latency experienced by data while they are elaborated at VNFs into hosts and while they traverse the VLs. Thus, the selection is not constrained by the hosts distance, but by the overall latency offered by the 5GT-MTP at both network and host levels. More specifically, this strategy seeks for the pair of hosts and for the VLs that minimize the accumulated processing latency at hosts and the network latency at the VLs, provided that bandwidth, CPU, memory and storage capacity demands are fulfilled and the overall latency constraint is honored. This strategy offers the lowest overall latency performance at the cost of spreading the resource utilization across both hosts and network links.

## V. PERFORMANCE EVALUATION

We now assess the performance of our solution by focusing on the cluster-based approach; the performance evaluation of the other schemes is omitted due to the lack of space. Our reference scenario is a fat-tree [15] with Level 1 abstraction (16 hosts in a fully-connected topology), and three services, each including between 5 and 10 VNFs.

We compare our strategy against a stochastic optimization approach, which is based on a genetic algorithm (GA), introduced in Sec. V.A.

*A. Our benchmark*

As a benchmark for our heuristic, we consider a genetic algorithm [13]. In GAs, a solution is represented as a *chromosome*, which is in turn composed of a number of *genes*, each encoding a specific property. In our case, a chromosome is a specific VNF placement solution, while a gene corresponds to a specific host, together with the set of VNF instances placed at it. Starting from a pool of initial chromosomes, which in our case contains $K$ random VNF-to-host assignments, a GA operates iteratively for a number of *generations* applying genetic operations to selected chromosomes to produce *offspring* (i.e., new chromosomes) of better quality according to a *fitness* function. The main genetic operations are *crossover* and *mutation*:

**Crossover**. At each generation, with rate rc, the genes of two chromosomes are combined to derive a new one. To improve the quality of the offspring, we introduce a specific gene-quality metric, and select the highest-quality genes of the two parents [14]. In other words, if we are minimizing cost, each gene is characterized by the sum of the costs of the VNFs placed at its host.

**Mutation**. With a very low probability (rm), each chromosome is subject to random changes to avoid being trapped into local optima. In our GA, this is implemented by randomly swapping VNFs between two genes.

At the end of each generation, a new solution pool is created by selecting the top-*K* chromosomes of the population according to a *fitness function*. A cost-minimizing algorithm uses as fitness functions the overall placement cost and latency, respectively, as defined in Section IV.A. The algorithm terminates by returning the chromosome with the highest fitness function value. Note that for a chromosome to be included in the pool, the capacity and delay constraints are always checked.

*B. Performance of the cluster-based approach*

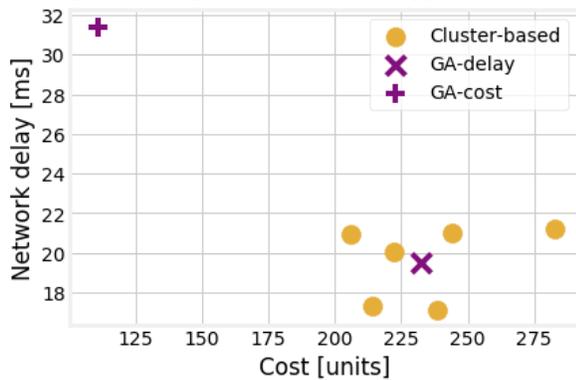

**Figure 3: Cost and delay yielded by the cluster-based approach for different number of clusters (yellow dots) and the benchmark (purple markers).**

Figure 3 shows the cost and delay associated with the clustering-based approach; each yellow dot therein corresponds to a specific number of clusters, varying from 1 to 7. It is easy to see that changing the number of clusters leads to different cost/delay trade-offs.

The two purple markers correspond to the results of the GA-based benchmark under the two objectives it supports: when it is set to minimize delay, the resulting configuration is similar to the one generated by the cluster-based approach. Setting the GA algorithm to minimize costs results in significantly lower costs than the cluster-based approach, but in a much higher delay.Conclusion

We addressed the relevant problem of designing a service orchestrator in 5G systems that efficiently supports vertical services while exploiting (if needed) services and resources made available by other administrative domains. We proposed a system architecture and discussed different levels of abstraction of physical resources that can be used at the orchestrator to make decisions. We then presented low-complexity algorithms that aim to minimize the network provider's cost and/or the service latency, while meeting the verticals' service requirements.

Beside extending our numerical performance evaluation, future work will be conducted mainly along the two following lines. First, resource orchestration is one of the components of the service instantiation or modification operations performed at the 5GT-SO. Further research is needed to devise efficient algorithms for the segmentation of NFV-NSs and mechanisms for service federation. Second, within the 5G-TRANSFORMER project we plan to realize a proof-of-concept of the proposed 5GT-SO architecture and resource orchestration algorithms, showing their scalability and efficacy in real-world situations.

ACKNOWLEDGMENTS

This work has been partially funded by the EU H2020 5G-Transformer Project (grant no. 761536).